\documentclass{svproc}

\usepackage{url}
\usepackage{graphicx}

\begin{document}
\bibliographystyle{unsrt}
\mainmatter              
\title{Thermonuclear Supernovae: Prospecting in the Age of Time-Domain and Multi-Wavelength Astronomy
}
\titlerunning{Thermonuclear Supernovae}  
%
\author{Peter Hoeflich\inst{1} \and Chris Ashall\inst{1} \and Alec Fisher\inst{1} \and Boyan Hristov\inst{1} \and David Collins \inst{1}  \and Eric Hsiao\inst{1} \and Ingo Wiedenhoever\inst{1} \and S. Chakraborty \inst{1}  \and Tiara Diamond \inst{2}}
\authorrunning{Peter Hoeflich et al.} 
%
%
\institute{Florida State University, Tallahassee, FL 32309, USA,
\email{phoeflich@fsu.edu} \and
NASA Goddard Space Flight Center, Greenbelt, MD 20771, USA}

\maketitle              

\begin{abstract}
 We show how new and upcoming advances 
in the age of time-domain and multi-wavelength astronomy will open up a new venue to probe the diversity of SN~Ia. 
We discuss this in the context of  the 
ELT (ESO), as well as space based instrument such as James Webb Space Telescope (JWST).
As examples we demonstrate how the power of very early observations, within hours to days after the explosion, and very late-time observations, such as light curves and mid-infraread spectra beyond 3 years,  can be used to probe the link to progenitors and explosion scenarios.
We identify the electron-capture cross sections of $Cr$, $Mn$, and $Ni/Co$ as one of the limiting factors we will face in the future.   
\end{abstract}

\noindent{\bf Introduction:}
Thermonuclear Supernovae,  stellar explosions of White Dwarf Stars (WD)/the degenerate C/O cores of low mass stars are important for understanding the Universe. As well as being one of the building blocks and drivers of modern cosmology, they are also important for understanding the origin of elements, 
and are laboratories for the explosion physics of WDs in close binary systems. Here, we focus on new developments.
For a general discussion from our perspective, see \cite{2017hsn..book.1151H,hoeflich06,hoeflich2013}.
Recently  advances in observations and theory have caused new problems to emerge.    
One of these is the discrepancy in the Hubble constant $H_o$ obtained using the Microwave background ($66.93 \pm 0.62 km/s/Mpc$,\cite{2016A&A...594A...1P})
and that obtained using the empirical SNe~Ia-based 
methods ($73.24 \pm 1.74$, \cite{2016ApJ...826...56R}).
This discrepancy may have direct 
consequences for: the interpretation of the Big-bang nucleosynthesis (Li-problem), high precision cosmology,   
early Black Hole formation, and new physics beyond the high-energy standard model.  
\begin{figure} 
\vskip -0.2cm
\begin{center}$
\begin{array}{c}
\includegraphics[width=1.00\textwidth]{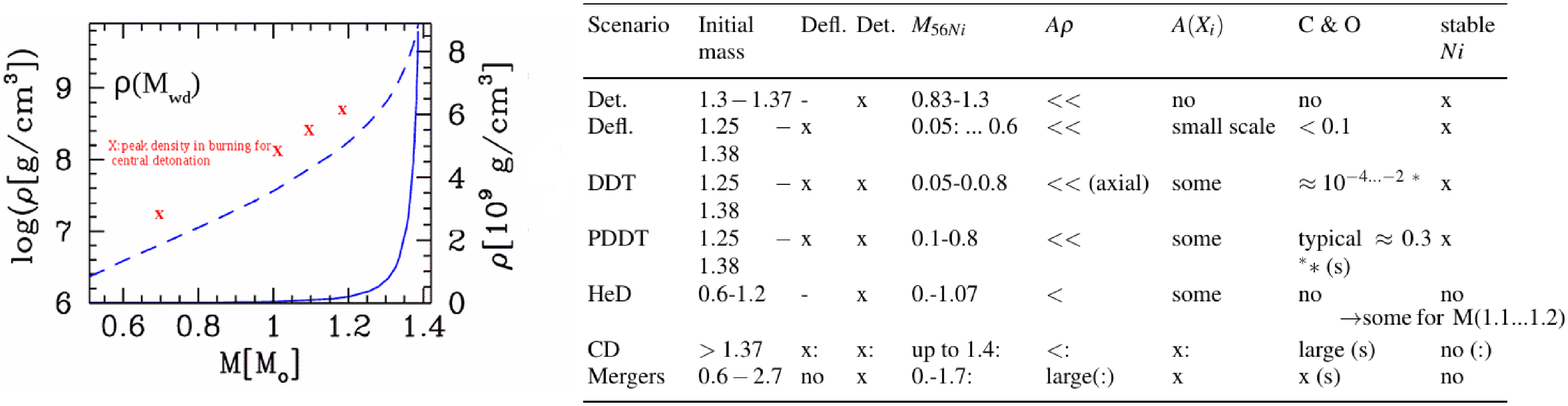} 
\end{array}$
\end{center}
\vskip -20pt
\caption{{\it Right:}
Characteristics of explosion scenarios including the range of total mass $M_{WD}$ in $M_\odot$, modes of burning, $M_{56Ni}$ production in $M_\odot$, asymmetries in 
density $A \rho$ and abundances $ A(X_i)$, presence of unburned $C/O$ 
and stable $Ni$. {\it x} denotes the presence of the feature.
 The production of electron capture isotopes depends 
on the nuclear physics, which is dominated by the density of burning,  vs. hydrodynamical time scales ($\approx 1 s$). The exact production
depends on the nuclear rates and recent revisions \cite{brach00,thielmann2018}.
The production of electron capture isotopes becomes important beyond $10^8 g/cm^3$.
{\it Left:} The central density of the WD is shown as a function of $M_{WD}$. For   
$M_{Ch}$ explosions this indicates the highest density of burning. In contrast, for HeDs, the detonation waves compress the material and increases the density. Therefore the corresponding relation for HeDs is indicated by red dots. E.g. to first order and as an upper limit, we may expect similar electron capture isotopes in HeD
and DDTs at 1.2 and 1.3 $M_\odot$, respectively. Note, however, that the  duration of compression by a detonation is shorter than the 
WD expansion time scale resulting in about 1/2 the shift in abundance with respect to EC isotopes \cite{hk96,hoeflich06}.
}
\label{fig1}
\begin{center}$
\begin{array}{c}
\includegraphics[width=0.90\textwidth]{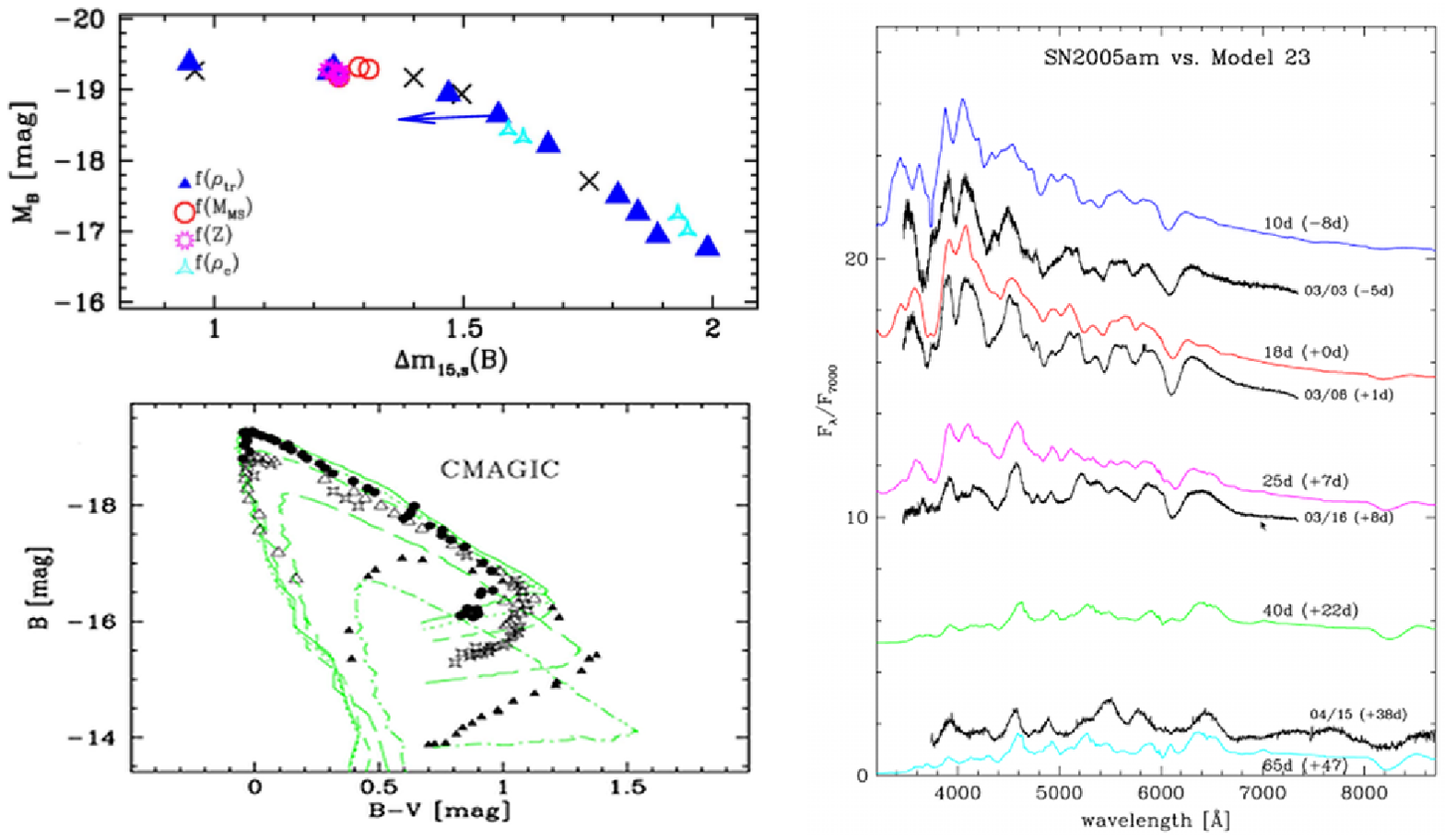} 
\end{array}$
\end{center}
\vskip -15pt
\caption{Comparison of observable properties of 'classical`, spherical delayed-detonation models with observations \cite{h17}.
We show the width-luminosity $\Delta m_{15,s} (B)$ ({\it upper left}) and the brightness-color relation of the normal bright
 SN2005M, SN2004eo, SN2005am and the subluminous SN2005ke (black X) and models (blue)
with transition densities of $27, ~23, ~16, 8 \times 10^6~g~cm^{-3}$ originating from a WD with a main sequence mass of $M_{MS}= 5 M_\odot$, an initial central density $\rho_c=2\time 10^9 g/cm^3$ and
solar metallicity $Z$ unless denoted.  
 As example, the comparison between the normal bright SN2005am and model spectra is shown on the right panel.
}
\label{fig2}
\end{figure}

\begin{figure}[ht] 
\begin{center}$
\begin{array}{c}
\includegraphics[width=0.95\textwidth]{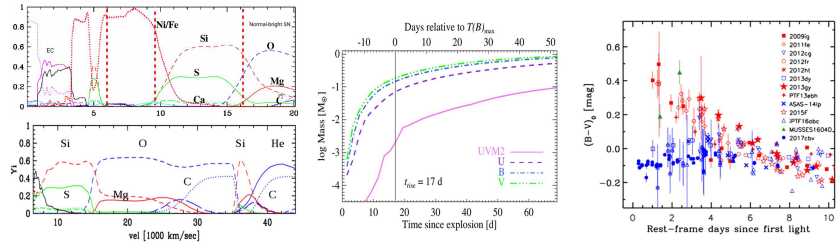}  
\end{array}$
\end{center}
\vskip -15pt
\caption{Nuclear signatures are a possible probe of the donor star for SD systems  and $M_{Ch}$ mass explosions.
 {\it  Upper left:} The overall abundance structure as a function of velocity $[in~1000 km/sec]$ of our reference model.
 {\it Lower left:} Same but the very outer layers for a DDT model with a He-star as donor.
{\it Middle:} The mass exposed in the UV, U, B and V bands \cite{Gall2018})as a function of time. {\it Right:} Observations right after the explosion, where the very outer layers are probed, indicate a bifurcation among objects with similar subsequent LC and spectra \cite{stritzinger18}. 
}                                                        
\label{fig3}
\end{figure}
\begin{figure*}
\begin{center}$
\begin{array}{cc}
\includegraphics[width=0.6\textwidth]{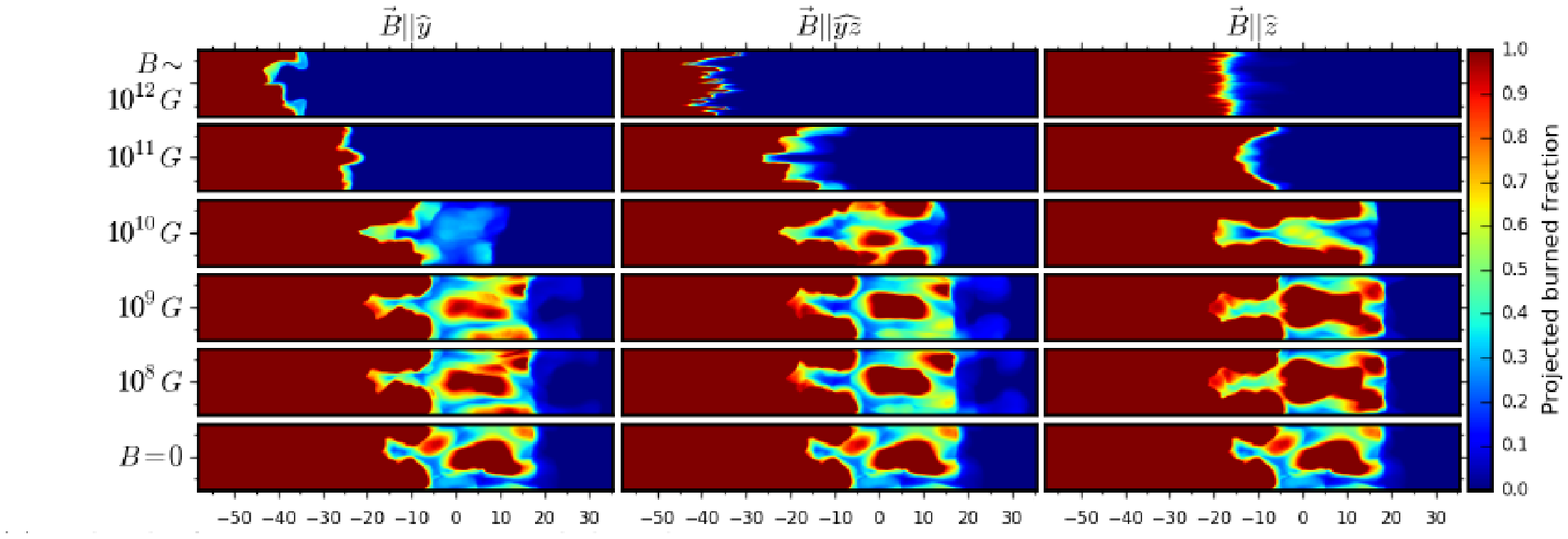} &
\includegraphics[width=0.4\textwidth]{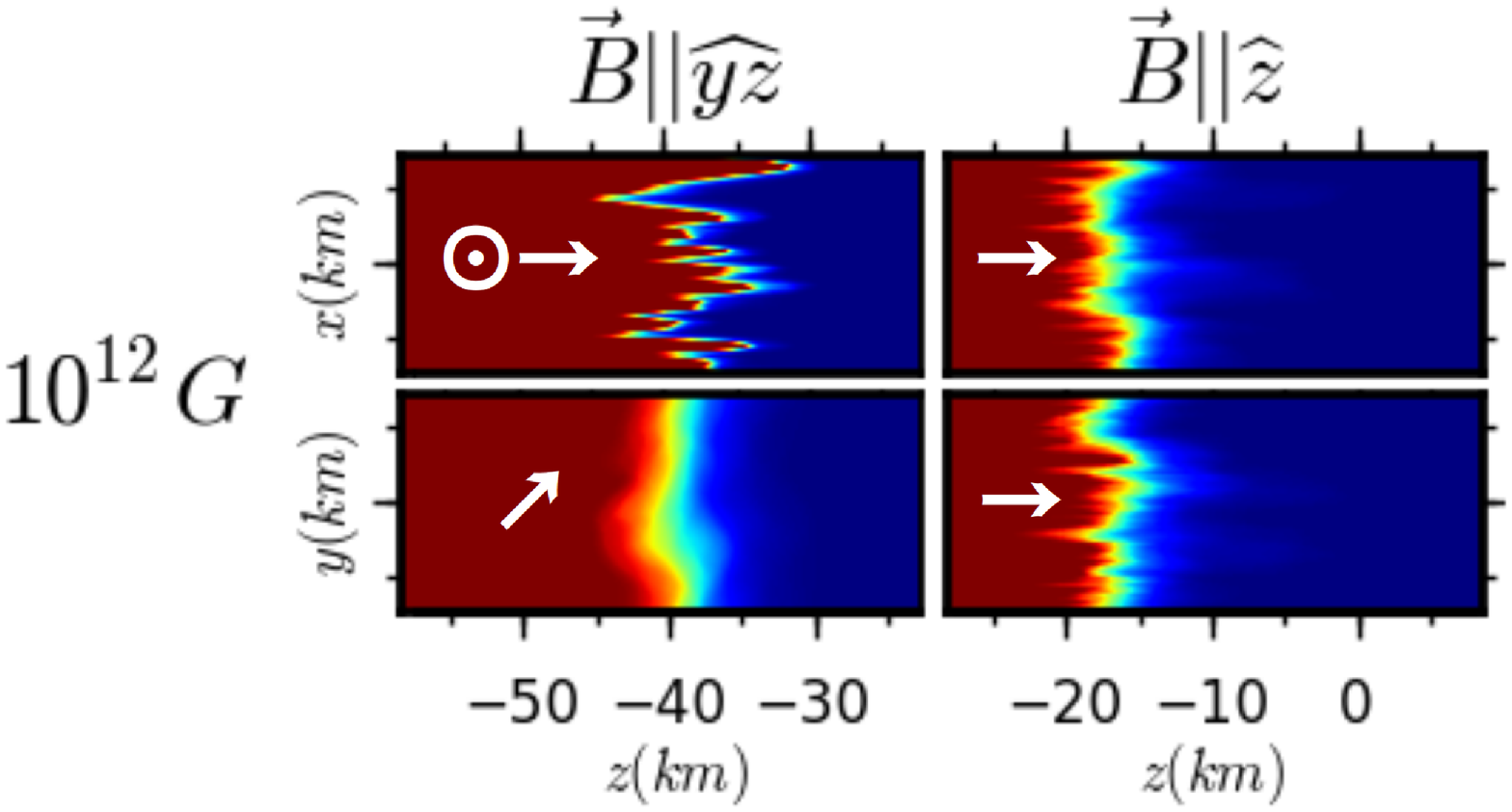} 
\end{array}$
\end{center}
\vskip -20pt
\caption{Possible consequence of multi-dimensional effects on the production of EC isotopes.
We show the influence of $B$ fields  on the normalized burned fraction nuclear burning front in a flux tube with properties
corresponding to the regime of non-distributed burning (based on Enzo). The axis are in km.
The  initial $B$-field is parallel ({\it left}), at $45^{\circ}$ 
({\it middle}), and orthogonal to the flux tube ({\it right}).
 For $B$ close to the saturation field ({\it right plot}),
comb-like structures occur which increase the surface and, thus, the burning reate by factors of 4-5.   
$B$ emerge as important component with $B > 10^{4...6} G$ both from
late-time NIR spectra and LCs \cite{h04,penney14,tiara15}, and may suppress excessive strong Rayleigh-Taylor mixing
which would degrade a tight $\Delta m_{15}$ relation.}
\label{fig4}
\vskip -15pt
\end{figure*}
\begin{figure} 
\begin{center}$
\begin{array}{c}
\includegraphics[width=0.85\textwidth]{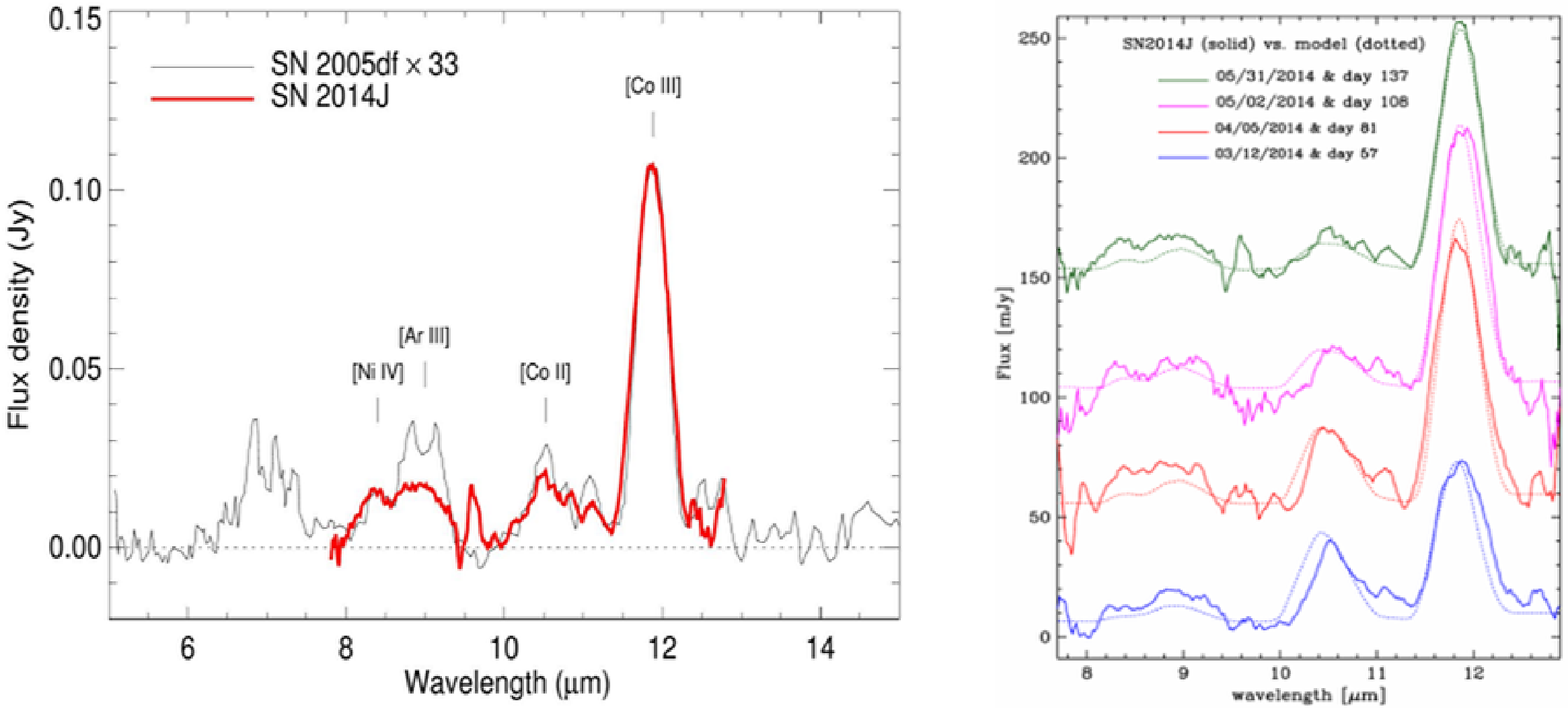}  
\end{array}$
\vskip -10pt
\caption{
{\it Left:} The MIR spectra at day $135 $ of SN~2005df the Spitzer Space Telescope (red, \cite{Gerardy}) and SN~2014J (black).
{\it Right:} The time series for SN2014J obtained at the Grand Canari Telescope (right) \cite{Telesco2015}. The dotted lines give the 
synthetic spectra of our reference DDT model with the wavelength in $\mu m$.
}
\label{fig5}       
\end{center}
\begin{center}$
\begin{array}{c}
\includegraphics[width=0.9\textwidth]{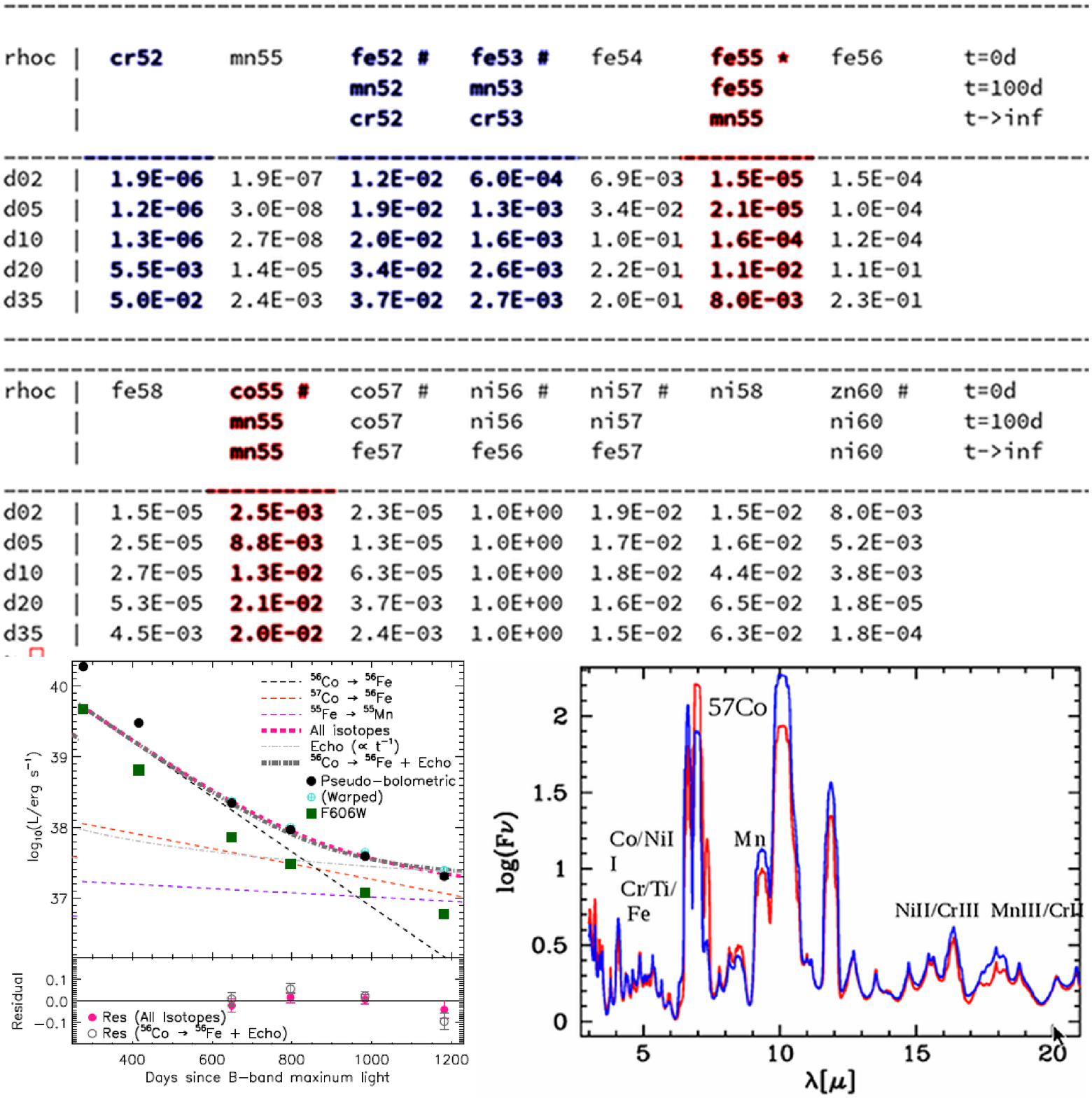} 
\end{array}$
\vskip -10pt
\caption{Ultra-late time observations are a probe for high density burning. {\it Top:}
EC isotopes at 0, 100 days and infinity for DDT models with $\rho_c$ between $2 ... 35 \times 10^8 g/cm^3$ (d02...d35)
for a delayed-detonation model with $\rho_c$ between $2 ... 35 \times 10^8 g/cm^3$ (d02...d35) for various times $t$).
{\it Lower left:} The bolometric and monochomatic LCs, obtained by the Hubble Space telescope, become dominated by rare, short
lived isotopes rather than the positron decay channel of $^{56}Co$  (from \cite{yang2018}), and 
require proton trapping which requires initial $B$ fields larger than $ 10^6 G$. {\it Lower right:}  
Predicted spectra at day 3000 are shown    
for $0 B$ (red) and $10^6 G$ (blue). The precence of  the electron capture isotopes $Cr$, $Mn$, and $Ni/Co$ should be noted.  
}
\label{fig6}
\end{center}
\end{figure}

The majority of Type Ia supernovae appear to be rather homogeneous with a well defined luminosity decline relation of light curves $\Delta m_{15}$ \cite{p93} and similar spectra. However,  there is in fact some diversity in their observations which has been hypothesized to be due to various progenitor channels and explosion scenarios.
Potential progenitor systems may either consist of two WDs, called a 
double degenerate (DD) system, or a single WD with a donor which may be main sequence, red giant, or Helium (${He}$) star, called a single degenerate (SD) system \cite{Stephano11}. The various explosion scenarios can be distinguished
by three possible triggering mechanisms: a) Compressional heat in a slow accretion triggers the explosion when the WD approaches (!)
the Chandrasekhar mass $M_{Ch}$ in either SD or DD systems. The flame propagates as a detonation (Det.), a deflagration (Defl.) or, more likely, starts as 
deflagration and transitions to a detonation with or without a pulsation phase (DDT,PDDT); b) Heat released on dynamical time scales  
triggers a detonation of a DD-system (dynamical mergers); c) in Helium detonations (HeDs,) a surface He-detonation triggers a detonation in a C/O core of a sub-$M_{Ch}$ WD with 
a He-star companion. Core-degenerates (CD) are explosions within a Red-Supergiant by a) or b). 

From theory, the empirical SNe~Ia relations $dm_{15}$ and $CMAGIC$ (Fig. 2) for cosmology are stable because basic nuclear 
physics determines: the structure of the progenitor WD, the explosion physics, and the average expansion velocities. This 'Stellar amnesia` 
leads to similar light curve shapes and spectral evolution.
All scenarios may contribute to the SNe~Ia population but there is observational evidence that one scenario dominates (Figs. \ref{fig1} \& \ref{fig2} )\cite{contr10,h17}.
However which one dominates is heavily discussed in the community. Dynamical mergers are not likely as they predict high  continuum polarization and aspherical explosions, this is not seen in the data \cite{patat12,rest05}.
The DDT ($M_{Ch}$) seems to explain most of the observed properties of SNe~Ia (Fig. \ref{fig2}), 
and DDT model-based, $\delta-Ceph.$-independent distances give an  $H_o = 68 \pm 4 km/Mpc/s$ \cite{hk96,h17}.
However, HeDs have recently become a serious contender because their main-flaw, the need for a large He-layers on the surface of the WD, can be migrated by 
a mixing of the He and C, as long as the $M_{WD}>1.1 M_{\odot}$ \cite{2014ApJ...797...46S}. In this scenario the optical spectra and LCs become
 similar to DDTs.
However, these  HeDs result in systematically larger $H_o$ when analyzing observations.   

\noindent{\bf New early time observations} \cite{stritzinger18}  provide a new tool to probe the outermost $10^{-4...-5}~ M_\odot$
 layers of the ejecta (Fig. \ref{fig3}).   
 Within the $M_{Ch}$ scenarios like DDTs these outermost layers consist of 
H- and He for main-sequence/red-giant and He-star donors, respectively. When the detonation burning passes through the 'surface'-layers 
 the burning time scales for hydrogen burning are too long \cite{HS09} but a He/C mixture would ignite leaving a nuclear signature 
at high velocity and during the first 2-5 days as observed (Fig. \ref{fig3}, lower left \& middle) . 
Alternative explanations may be high mass HeDs, PDDTs or ongoing interaction pending 
further analysis.

\noindent{\bf Late and ultra-time observations in the mid-infrared} are a 
novel tool to probe for the distribution electron capture isotopes and, with it, the underlying physics of flames (\ref{fig4}).
The detection of stable ${Ni}$ at late times can distinguish between $M_{Ch}$ explosions or high mass ($M\approx 1.2 M_\odot$ HeDs 
from mergers  or low mass HeDs (Fig \ref{fig5}).
Unfortunately, the $^{57}Ni/Co$ production, which is analyzed in very late time light curves of SNe Ia, is rather insensitive to density of burning for $M_{WD} $ larger than $1.1 M_\odot$. 
This does not allow us to distinguish massive HeD from $M_{Ch}$ models.
However, as the density of burning increases the $Cr, ~Mn $ and stable $Ni$ abundances becomes more prominent, see the table in  Fig \ref{fig6}. 
These lines are predominantly located in the  the mid-IR and will be observable with JWST in the near future. By doppler shifts, line profiles will 
provide the spatial location and test for mixing vs. nuclear effects. 
JWST and ELT will open up the parameter space
for many SNe~Ia and allow to use the spectra during the $^{57}Co$ regime (Fig \ref{fig6}) emphasizing the need for high-precision,
electron capture nuclear data for $Mn,Cr,Ni$ isotopes on the 20 to 40 \% level based on our tests.

\noindent{{\bf Acknowledgements:} We thank the FSU Foundation for funds which allowed to attend the meeting. The work has been 
supported by the NSF grant 1715133.}
\bibliography{article}

\begin{thebibliography}{10}

\bibitem{2017hsn..book.1151H}
P.~{Hoeflich}.
\newblock {\em {Explosion Physics of Thermonuclear Supernovae and Their
  Signatures}}, page 1151.
\newblock 2017.

\bibitem{hoeflich06}
P.~{Hoeflich}.
\newblock {Physics of type Ia supernovae}.
\newblock {\em Nuclear Physics A}, 777:579--600, 2006.

\bibitem{hoeflich2013}
P.~{Hoeflich}, P.~{Dragulin}, J.~{Mitchell}, B.~{Penney}, B.~{Sadler},
  T.~{Diamond}, and C.~{Gerardy}.
\newblock Properties of SN~Ia progenitors from light curves and spectra.
\newblock {\em Frontiers of Physics}, 8:144--167, April 2013.

\bibitem{2016A&A...594A...1P}
{Planck Collaboration}, R.~{Adam}, P.~A.~R. {Ade}, N.~{Aghanim}, and {et.al.}
\newblock {Planck 2015 results. I. Overview of products and scientific
  results}.
\newblock {\em AA}, 594:A1, September 2016.

\bibitem{2016ApJ...826...56R}
A.~G. {Riess}, L.~M. {Macri}, S.~L. {Hoffmann}, and {et.al.}
\newblock {A 2.4\% Determination of the Local Value of the Hubble Constant}.
\newblock {\em ApJ}, 826:56, July 2016.

\bibitem{brach00}
F.~Brachwitz, D.~Dean, W.~R. Hix, K.~Iwamoto, K.~Langanke, G.~Martinez-Pinedo,
  K.~Nomoto, M.~Strayer, F.-K. Thielemann, and H.~Umeda.
\newblock The role of electron captures in chandrasekhar mass models for type
  ia supernovae.
\newblock {\em ApJ}, 536:934, 2000.

\bibitem{thielmann2018}
F.-K. {Thielemann}, J.~{Isern}, A.~{Perego}, and P.~{von Ballmoos}.
\newblock {Nucleosynthesis in Supernovae}.
\newblock {\em Space Science Review}, 214:62, April 2018.

\bibitem{hk96}
P.~{Hoeflich} and A.~{Khokhlov}.
\newblock {Explosion Models for Type IA Supernovae: A Comparison with Observed
  Light Curves, Distances, H 0, and Q 0}.
\newblock {\em ApJ}, 457:500, 1996.

\bibitem{h17}
P.~{Hoeflich}, E.~Y. {Hsiao}, C.~R. {Ashall}, and {the CSP collaboration}.
\newblock {Light and Color Curve Properties of Type Ia Supernovae: Theory
  Versus Observations}.
\newblock {\em ApJ}, 846:58, September 2017.

\bibitem{Gall2018}
C.~{Gall}, M.~D. {Stritzinger}, C.~{Ashall}, E.~{Baron}, C.~R. {Burns},
  P.~{Hoeflich}, and {et.al.}
\newblock {Two transitional type Ia supernovae located in the Fornax cluster
  member NGC 1404: SN 2007on and SN 2011iv}.
\newblock {\em ApJL}, 35.

\bibitem{stritzinger18}
M.~D. {Stritzinger}, B.~J. {Shappee}, A.~L. {Piro}, C.~{eg.al.},
\newblock {Red vs Blue: Early observations of thermonuclear supernovae reveal
  two distinct populations?}
\newblock {\em ApJ}, 864, L35

\bibitem{h04}
P.~{Hoeflich}, C.~L. {Gerardy}, K.~{Nomoto}, K.~{Motohara}, R.~A. {Fesen},
  K.~{Maeda}, T.~{Ohkubo}, and N.~{Tominaga}.
\newblock {Signature of Electron Capture in Iron-rich Ejecta of SN 2003du}.
\newblock {\em ApJ}, 617:1258--1266, 2004.

\bibitem{penney14}
R.~{Penney} and P.~{Hoeflich}.
\newblock {Thermonuclear Supernovae: Probing Magnetic Fields by Positrons and
  Late-time IR Line Profiles}.
\newblock {\em ApJ}, 795:84, November 2014.

\bibitem{tiara15}
T.~R. {Diamond}, P.~{Hoeflich}, and C.~L. {Gerardy}.
\newblock {Late-time Near-infrared Observations of SN 2005df}.
\newblock {\em ApJ}, 806:107, June 2015.

\bibitem{Gerardy}
C.~L. {Gerardy}, P.~{Hoeflich}, R.~A. {Fesen}, and {et.al.}
\newblock {SN 2003du: Signatures of the Circumstellar Environment in a Normal
  SNe~Ia?}
\newblock {\em ApJ}, 607:391--405, 2004.

\bibitem{Telesco2015}
C.~M. {Telesco}, P.~{H{oe}flich}, D.~{Li}, and {et.al.}
\newblock {Mid-IR Spectra of Type Ia SN 2014J in M82 Spanning the First 4
  Months}.
\newblock {\em ApJ}, 798:93, January 2015.

\bibitem{yang2018}
Y.~{Yang}, L.~{Wang}, D.~{Baade}, P.~J. {Brown}, A.~{Cikota}, M.~{Cracraft},
  P.~A. {H{\"o}flich}, J.~R. {Maund}, F.~{Patat}, W.~B. {Sparks},
  J.~{Spyromilio}, H.~F. {Stevance}, X.~{Wang}, and J.~C. {Wheeler}.
\newblock {Late-time Flattening of Type Ia Supernova Light Curves: Constraints
  from SN 2014J in M82}.
\newblock {\em ApJ}, 852:89, January 2018.

\bibitem{p93}
M.~M. {Phillips}.
\newblock {The absolute magnitudes of Type IA supernovae}.
\newblock {\em Astrophysical Journal, Letters}, 413:L105--L108, 1993.

\bibitem{Stephano11}
R.~{Di Stefano}, R.~{Voss}, and J.~S.~W. {Claeys}.
\newblock {Single- And Double-degenerate Models Of Type Ia Sne, Nuclear-burning
  White Dwarfs, Spin, And Supersoft X-ray Sources}.
\newblock In {\em AAS/High Energy Astrophysics Division}, volume~12 of {\em
  AAS/High Energy Astrophysics Division}, page 33.04, 2011.

\bibitem{contr10}
C.~{Contreras}, M.~{Hamuy}, {Phillips}, and {et.al.}
\newblock {The Carnegie Supernova Project: First Photometry Data Release}.
\newblock {\em Astronomical Journal}, 139:519--539, 2010.

\bibitem{patat12}
F.~{Patat}, P.~{Hoeflich}, D.~{Baade}, and {et.al.}
\newblock {VLT Spectropolarimetry of the Type Ia SN 2005ke. A step towards
  understanding subluminous events}.
\newblock {\em AA}, 545:A7, 2012.

\bibitem{rest05}
A.~{Rest}, N.~B. {Suntzeff}, K.~{Olsen}, and {et.al.}
\newblock {Light echoes from ancient supernovae in the Large Magellanic Cloud}.
\newblock {\em Nature}, 438:1132--1134, 2005.

\bibitem{2014ApJ...797...46S}
K.~J. {Shen} and K.~{Moore}.
\newblock {The Initiation and Propagation of Helium Detonations in White Dwarf
  Envelopes}.
\newblock {\em ApJ}, 797:46, December 2014.

\bibitem{HS09}
P.~{H{\"o}flich} and B.~E. {Schaefer}.
\newblock {X-ray and Gamma-ray Flashes from Type Ia Supernovae?}
\newblock {\em ApJ}, 705:483--495, November 2009.

\end{thebibliography}
\end{document}